\documentclass[a4paper, 11pt]{article}
\usepackage{mathrsfs}
\usepackage{hyperref}
\usepackage{color}
\usepackage{cite}
\usepackage{subfigure}
\usepackage{caption}
\usepackage[pdftex, a4paper]{geometry}
\usepackage{fancyhdr}
\usepackage{paralist}
\usepackage{graphics}
\usepackage{amsfonts}
\usepackage{dsfont}
\usepackage{mathrsfs}
\usepackage{amssymb}
\usepackage{bbm}
\usepackage{epsfig}
\usepackage{amsmath}
\usepackage{appendix}
\usepackage[english]{babel}
\usepackage[all]{xy}

\newtheorem{theorem}{Theorem}
\newtheorem{definition}{Definition}

\newtheorem{lemma}{Lemma}

\newtheorem{proposition}{Proposition}
\newtheorem{example}{Example}

\begin{document}
\title{\bf An Improved Stability Condition
for Kalman Filtering with Bounded Markovian Packet Losses}
\date{}

\author{Junfeng Wu, Ling Shi, Lihua Xie, and Karl Henrik Johansson}
\maketitle

\begin{abstract}
In this paper, we consider the peak-covariance stability of Kalman filtering subject to packet losses.
The length of consecutive packet losses is governed by a time-homogeneous
finite-state Markov chain.
We establish a sufficient condition for peak-covariance stability and show that this stability check can be recast as a linear matrix inequality (LMI) feasibility problem.
Comparing with the literature, the stability condition given in this paper is invariant with respect to similarity state  transformations; moreover, our condition is proved to be less conservative than the existing results.
Numerical examples
are provided to demonstrate the effectiveness of our result.
\end{abstract}


\section{Introduction}

Networked control systems are closed-loop systems, wherein sensors, controllers and actuators are interconnected through a communication network. In the last decade, advances of modern control, micro-electronics, wireless
communication and networking technologies
have given birth to a considerable number of networked control applications.

In networked control systems, state estimation such as using a Kalman filter is necessary whenever precise measurement of the system state cannot be obtained.
When a Kalman filter is running subject to intermittent
observations, the stability of the estimation
error is affected by not only the system dynamics
but also by the statistics of the packet loss process. The stability of Kalman filtering with packet drops has been intensively studied in the literature.
In~\cite{Sinopoli2004,Plarre09tac,shi-tac10,yilin10bound,kar2012kalman}, an independently and identically distributed (i.i.d.) Bernoulli packet loss has been considered.
Some other research works assume the packet drops are due to the Gilbert-Elliott channel~\cite{gilbert1960capacity,elliott1963estimates},
which are governed by a time-homogeneous Markov chain. {Huang and Dey}~\cite{huang-dey-stability-kf} introduced the notion of peak covariance, which describes the upper envelope of the sequence of error covariance matrices for the case of an unstable scalar system.
They focused on its stability with
Markovian packet losses and gave a sufficient stability condition. The stability condition was further improved  in~\cite{xie2007peak,xie2008stability}.
In~\cite{DBLP:journals/corr/WuSAJ14}, the authors
proved that the peak-covariance stability implies mean-square
stability for general random packet drop processes, if the system matrix has no defective eigenvalues on the unit circle.
In addition to the peak-covariance stability,
the mean-square stability was considered for some classes of linear systems in~\cite{you2011mean,yilin12criticalvalue}, and weak  convergence of the estimation error covariance was studied in~\cite{xie2012stochastic}.

In the aforementioned packet loss models, the length of consecutive packet losses can be infinitely large. In contrast, some works
also consider bounded packet loss model, whereby the length of
consecutive packet losses is restricted to be less than a finite integer.
A real example of bounded packet losses is the WirelessHART (Wireless
Highway Addressable Remote Transducer) protocol, which is the state-of-the-art
 wireless communication solution for process automation applications.
In WirelessHART, there are two types of time slots: one is the dedicated time slot allocated to a specific field device for
time-division multiple-access (TDMA) based transmission and
the other is the shared time slot allowing contention-based transmission.
A contiguous group of time slots during a constant period of time
forms a superframe, within which every node is guaranteed at least one time slot for data
communication.
Various networked control problems with bounded packet loss model have been  studied, e.g.,~\cite{xiong2007stabilization,wu2007design}; while the stability of Kalman filtering with the bounded packet loss model was rarely discussed. In~\cite{xiao2009kalman}, the authors gave a first attempt to the stability issue related to the Kalman filtering with bounded Markovian losses.
They provided a sufficient condition for peak-covariance stability, the stability notion studied in~\cite{huang-dey-stability-kf,xie2007peak,xie2008stability}.
Their result has established a connection between peak-covariance stability and the dynamics of the underlying system and the probability transition matrix of the underlying packet-loss process. In this paper, we consider the same problem as in~\cite{xiao2009kalman} and improve the condition thereof.
The main contributions of this work are summarized as follows:
\begin{enumerate}
\item We present a sufficient condition for peak-covariance stability of
the Kalman filtering subjected to bounded Markovian packet losses. Different from that of~\cite{xiao2009kalman}, this stability check can be recast as
a linear matrix inequality (LMI) feasibility problem.
\item We compare the proposed condition with that of~\cite{xiao2009kalman}. We show both theoretically and numerically that the proposed stability condition is invariant  with respect to similarity state transformations, while the one given in~\cite{xiao2009kalman}
may generate opposite conclusions under different similarity transformations. Moreover, the analysis also suggests that our condition is less conservative than the former one.
\end{enumerate}
The remainder of the paper is organized as follows. Section~\ref{section:problem-setup} presents the mathematical models of the system and packet losses,
and introduces the preliminaries of Kalman filtering. Section~\ref{section:main-result}
provides the main results. Comparison with~\cite{xiao2009kalman} and numerical examples are
presented in Section~\ref{section:comparison}.
Some concluding remarks are drawn in the end.

\textit{Notations}: $\mathbb{N}$  is
the set of positive integers and $\mathbb{C}$ is the set of
of complex numbers. $\mathbb{S}_{+}^{n}$ is the set of $n$ by $n$  positive semi-definite matrices over the field $\mathbb{C}$. For a matrix $X\in \mathbb {C}^{n\times n}$,
$\sigma(X)$ denotes the spectrum of $X$, i.e., $\sigma(X)=\{\lambda:\det(\lambda I-X)=0\}$, and $\rho(X)$ denotes the spectrum radius of $X$.
$\|X\|$ means the $L_2$-norm on $\mathbb{C}^n$ or the matrix norm induced by $L_2$-norm. The symbol $\otimes$ represents the Kronecker product operator of two matrices.
For any matrices $A,~B,~C$ with compatible dimensions,
we have
$\mathrm{vec}(ABC)=(C'\otimes A)\mathrm{vec}(B)$,
where $\mathrm{vec}(\cdot)$ is the vectorization of
a matrix.
Moreover, the indicator function of a subset $\mathcal{A}\subset\Omega$ is a function
$\mathbf{1}_\mathcal{A}:~ \Omega \rightarrow \{ 0,1 \}$ where
$\mathbf{1}_\mathcal{A}(\omega)=1$ if $\omega\in \mathcal{A}$, otherwise
$\mathbf{1}_\mathcal{A}(\omega)=0$.


\section{Problem Setup}\label{section:problem-setup}
\subsection{System Model}
Consider the following discrete-time LTI system:
\begin{eqnarray}
x_{k+1} & = & Ax_k + w_k, \label{eqn:process-dynamics} \\
y_k & = & Cx_k + v_k, \label{eqn:measurement-equation}
\end{eqnarray} where $A\in \mathbb{R}^{n\times n}$ and $C\in \mathbb{R}^{m\times n}$, $x_k \in \mathbb{R}^{n}$ is the process state vector, $y_k \in \mathbb{R}^{m}$ is the observation vector,
$w_{k} \in \mathbb{R}^{n} $ and
$v_k \in \mathbb{R}^{m}$ are zero-mean Gaussian random vectors
with $\mathbb{E}[w_{k}w_{j}{'}]=\delta_{kj}Q~(Q\geq 0)$,
$\mathbb{E}[v_{k}v_{j}{'}] = \delta_{kj}R~(R > 0)$,
$\mathbb{E}[w_{k}v_{j}{'}] = 0 \; \forall j,k$. Note that $\delta_{kj}$ is the Kronecker
delta function with $\delta_{kj} = 1$ if $k=j$ and $\delta_{kj}=0$ otherwise.
The initial state $x_0$ is a zero-mean Gaussian random vector that is uncorrelated with $w_k$ and $v_k$. Its covariance is $\Sigma_0\geq 0$.
It can be seen that, by applying a similarity transformation, the unstable and stable modes of the LTI system can be decoupled. An open-loop prediction of the stable mode always has a bounded estimation error covariance, therefore, this mode does not play any key role in the problem considered below.
Without loss of generality, all eigenvalues of $A$ are assumed to have magnitudes not less than 1. We also assume that $(A,C)$ is observable and $(A,Q^{{1}/{2}})$ is controllable. We introduce the definition of the observability index of $(A,C)$, which is taken from~\cite{antsaklis2006linear}.
\begin{definition}\label{def:observability-index}
The observability index $\mathsf{I_o}$ is defined as the
smallest integer such that \\
$[C',A'C',\ldots,(A^{{I}_o-1})'C']'$ has rank $n$.
If $\mathsf{I_o}=1$, the system $(A,C)$ is called one-step observable.
\end{definition}

\subsection{Bounded Markovian Packet-loss Process}
In this paper, we consider the estimation scheme, where the raw measurements of the sensor $\{y_k\}_{k\in\mathbb{N}}$
are transmitted to the estimator over an erasure communication channel:
packets may be randomly dropped or successively received by the estimator.
Denote by a random variable $\gamma_k\in\{0,1\}$ whether or not $y_k$ is received at time $k$. If $\gamma_k=1$, it indicates that $y_k$ arrives error-free at the estimator; otherwise $\gamma_k=0$.
Whether $\gamma_k$ equals $0$ or $1$ is assumed to have been known
by the estimator before time $k+1$.
In order to introduce
the packet loss model, we further define a sequence of stopping times based
on $\{\gamma_k\}_{k\in\mathbb{N}}$, which presents the time instants at which packets are received by the estimator:
\begin{eqnarray}\label{def:stopping-times}
t_1&\triangleq& \min\{k:k\in\mathbb{N}, \gamma_k=1\},\notag\\
t_2&\triangleq& \min\{k:k>t_1, \gamma_k=1\},\notag\\
&\vdots&\\
t_j&\triangleq& \min\{k:k>k_{j-1}, \gamma_k=1\},
\end{eqnarray}
where we assume $t_0=0$ by convention. Then at the $j$th time instant
 the estimator successfully receives a measurement from the sensor.
The packet-loss process, $\tau_j$, is defined as
\begin{eqnarray*}
\tau_j&\triangleq& t_j-t_{j-1}-1.
\end{eqnarray*}
As for the model of packet losses, we assume that
the packet-loss process $\{\tau_j\}_{j\in\mathbb{N}}$ is modeled by a time-homogeneous ergodic Markov chain, where
$\mathcal{S}=\{0,\ldots,s\}$ is the finite-state space of the Markov chain with $s$ being the maximum length of consecutive lost packets allowed.
Here this Markov chain is characterized by a known transition probability matrix
$\mathbf{\Pi}\triangleq [\pi_{ij}]_{i,j\in\mathcal{S}}$ in which
\begin{equation}\label{eqn:transition-matrix}
\pi_{ij}\triangleq \mathbb{P}(\tau_{k+1}=j|\tau_{k+1}=i)\geq 0.
\end{equation}
Denote by $\pi\triangleq[\pi_0,\ldots,\pi_s]$
the stationary distribution of $\{\tau_j\}_{j\in\mathbb{N}}$.
Without loss of generality,
we assume that the initial distribution is $\mathbb{P}(\tau_{1}=j)=\pi_j$
and other cases can be treated in the same manner.

\subsection{Kalman Filtering with Packet Losses}
Sinopoli~{et~al.}~\cite{Sinopoli2004} shows that, when performed with intermittent observations, the optimal linear estimator is a modified Kalman filter. The modified Kalman filter is slightly different from
the standard one in that only time update is performed
when the data packet is lost at that time.
Define the minimum mean-squared error estimate and the one-step prediction
at the estimator respectively as
$\hat{x}_{k|k}\triangleq\mathbb{E}[x_k|\gamma_1y_1,\ldots,\gamma_ky_k]$
and
$\hat{x}_{k+1|k}\triangleq\mathbb{E}[x_{k+1}|\gamma_1y_1,\ldots,\gamma_ky_k]$.
 Let $P_{k|k}$ and $P_{k+1|k}$ be the corresponding estimation and prediction error covariance matrices, i.e.,
\begin{eqnarray*}
P_{k|k}&\triangleq&\mathbb{E}[(x_k-\hat x_{k|k}) (\cdot)'|\gamma_1y_1,\ldots,\gamma_ky_k]\\
P_{k+1|k}&\triangleq&\mathbb{E}[(x_{k+1}-\hat x_{k+1|k}) (\cdot)'|\gamma_1y_1,\ldots,\gamma_ky_k].
\end{eqnarray*}
These parameters can be computed recursively by a modified Kalman filter
(see~\cite{Sinopoli2004} for more details). In particular,
\begin{eqnarray}\label{eqn:DARE}
P_{k+1|k}&=&AP_{k|k-1}A'+Q\\
&&-\gamma_k AP_{k|k-1}C'(CP_{k|k-1}C'+R)^{-1}CP_{k|k-1}A'.\notag
\end{eqnarray}
To simplify notations,
we  denote
$P_k\triangleq P_{k|k-1}$ for shorthand
and define the functions $h$, $g$, $h^k$ and $g^k$: $\mathbb{S}^n_+ \to \mathbb{S}^n_+$ as
follows:
\begin{eqnarray}
h(X)&\triangleq& AXA'+Q,\label{eqn:h-func}\\
g(X)&\triangleq &AXA'+Q-AXC'
{(CXC'+R)^{-1}CXA'},\label{eqn:g-func}
\end{eqnarray}\vspace{2mm}
$h^k(X)\triangleq \underbrace{h\circ h \circ \cdots \circ h}_{k \hbox{~times}}(X)$ and
$g^k(X)\triangleq \underbrace{g\circ g \circ \cdots \circ g}_{k \hbox{~times}}(X)$,
where $\circ$ denotes the function composition.

\subsection{Problems of Interest}
To study the stability of Kalman filtering with packet losses,
one way is to study the asymptotic behavior of the expected prediction error covariance sequence. In the following we introduce the concept of peak-covariance stability, which is first studied in~\cite{huang-dey-stability-kf}.
To this end, we need the following auxiliary definitions, which are also introduced in~\cite{huang-dey-stability-kf},
\begin{eqnarray}
\alpha_1&\triangleq& \min\{k:k\in\mathbb{N}, \gamma_k=0\},\notag\\
\beta_1&\triangleq& \min\{k: k>\alpha_1, \gamma_k=1\},\label{def:alpha-j}\\
&\vdots&\notag\\
\alpha_j&\triangleq& \min\{k: k>\beta_{j-1}, \gamma_k=0\},\notag\\
\beta_j&\triangleq& \min\{k: k>\alpha_{j}, \gamma_k=1\},\label{def:beta-j}
\end{eqnarray}
where $\beta_0=0$ by convention.
It is straightforward to verify that $\{\alpha_j\}_{j\in\mathbb{N}}$
and $\{\beta_j\}_{j\in\mathbb{N}}$ are two sequence of stopping times (cf.,~\cite{durrett2010probability}).
\begin{definition}\label{def:peak-cov-stability}
The Kalman filtering system with packet losses is said to be peak-covariance stable if $\sup_{j\in\mathbb{N}} \mathbb{E} \| P_{\beta_j}\|<\infty$.
\end{definition}
Note that $\mathbb{E}\|P_{\beta_j+1}\|$, the mean of one-step prediction error
covariance at stopping time $\beta_j$, reflects the stability of Kalman filtering at packet reception times.
In the literature, stability of Kalman filtering with binary Markovian packet losses (driven by a two-state Gilbert-Elliott packet loss model)~\cite{huang-dey-stability-kf,xie2008stability,you2011mean}
and with i.i.d. packet losses~\cite{Sinopoli2004,shi-tac10} has been
intensively studied.
The main problem of this work is to study stability of
Kalman filtering with bounded Markovian packet-loss process.
As the packet loss is modelled differently, the stability
also behaves differently. Due to the nonlinearity of the Kalman filter, it seems
challenging to find necessary and sufficient stability conditions
for a general LTI system.
In Section~\ref{section:main-result}, we manage to give a sufficient peak-covariance stability
condition for general LTI systems with bounded Markovian
packet-loss process. Our result is mainly built on the prior work~\cite{xiao2009kalman}.
Compared with the result thereof, ours prevails
from at least two aspects. We will discuss in details
later in Section~\ref{section:comparison}.

\section{Main Result}\label{section:main-result}

In the following theorem, we will present a sufficient condition
for peak-covariance stability of Kalman filtering with bounded Markovian
packet-loss process.
\begin{theorem}\label{thm:main-thm}
Consider the system described in~\eqref{eqn:process-dynamics} and~\eqref{eqn:measurement-equation}, and the bounded Markovian
packet-loss process described by a probability transition matrix $\mathbf{\Pi}$ in~\eqref{eqn:transition-matrix}.
If there exists $K\triangleq[K^{(1)},\ldots,K^{(\mathsf{I_o}-1)}]$, where
$K^{(i)}$'s are matrices with compatible dimensions, such that
$\rho({H_K})<1$, where
\begin{equation}\label{def:H-k}
H_K\triangleq\mathrm{diag}\Big(A\otimes A,\ldots,(A\otimes A)^s\Big)
\left[\mathbf{P}'\otimes \mathbf{H}
+\mathbf{Q}'\otimes \mathbf{K}\right],
\end{equation}
and
$$\mathbf{H}\triangleq(\overline{A+K^{(1)}C})\otimes
(A+K^{(1)}C),$$
$$\mathbf{K}\triangleq
\sum_{l=2}^{\mathsf{I_o}-1}(\pi_{00})^{l-2}(\overline{A^l+K^{(l)}C^{(l)}})\otimes
(A^l+K^{(l)}C^{(l)}),$$
$\mathbf{P}\triangleq [\pi_{ij}]_{i,j\in\mathcal{S}/\{0\}}$
and $\mathbf{Q}\triangleq [\pi_{i0}\pi_{0j}]_{i,j\in\mathcal{S}/\{0\}}$;
then the state estimator is peak-covariance stable, i.e., $\sup_{j\in\mathbb{N}} \mathbb{E} \| P_{\beta_j}\|<\infty$.
\end{theorem}
Before proceeding to the proof, we first present  a few supporting definitions and lemmas.

Consider $k$ compositions of $g$ together. We introduce the following lemma.
\begin{lemma}\label{lemma:gx}
Consider the operator
\begin{eqnarray*}\phi_i({K}^{(i)},P)&\triangleq&(A^i+K^{(i)}C^{(i)})X(\cdot)^*+
[A^{(i)}~K^{(i)}]\left[\begin{array}{ccc}
Q^{(i)}& Q^{(i)}(D^{(i)})'\\
D^{(i)}(Q^{(i)})& D^{(i)}(Q^{(i)})(D^{(i)})'+R^{(i)}
\end{array}\right][A^{(i)}~K^{(i)}]^*,\\
&&~\forall i\in \mathbb{N},
\end{eqnarray*}
where $C^{(i)}\triangleq[C',~A'C',\cdots, (A')^{i-1}C']'$, $A^{(i)}\triangleq[A^{i-1},\cdots,A,~I]$,
$D^{(i)}=0$ for $i=1$ otherwise $D^{(i)}\triangleq\left[\begin{array}{cccc}
0& 0 & \cdots & 0\\
C& 0 & \cdots & 0\\
\vdots &\vdots& \ddots& \vdots\\
CA^{i-2}& CA^{i-3}& \cdots & 0\end{array}\right]$,
$Q^{(i)}\triangleq\mathrm{diag}(\underbrace{Q,\cdots,Q}_i)$,
$R^{(i)}\triangleq\mathrm{diag}(\underbrace{R,\cdots,R}_i)$,
and $K^{(i)}$ are of compatible dimensions.
 For any $X\geq 0$ and $K^{(i)}$, the following statement always holds
$$g^i(X)=\min_{K^{(i)}}\phi_i(K^{(i)},X)\leq \phi_i(K^{(i)},X).$$
\end{lemma}
{\it Proof.}
The result is readily established
when setting $B=I$ in Lemmas 2 and 3 in~\cite{xiao2009kalman}.
For $i=1$, The result is well known as Lemma 1 in~\cite{Sinopoli2004}.
\hfill$\square$

The following lemma is about the nonlinearity of $g$ operator: for $k\geq \mathsf{I_o}+1$, $g^k(X)$ is uniformly bounded no matter what the postive
semidefine matrix $X$ is.
\begin{lemma}[Lemma 5 in~\cite{huang-dey-stability-kf}]
\label{lemma:riccati>Io}
Assume that $(A,C)$ is observable and $(A,Q^{{1}/{2}})$ is controllable.
Define
$$
\mathbb{S}_0^n\triangleq\{P: 0\leq P\leq AP_0A'+Q,\hbox{~for some~} P_0\geq 0\},
$$
Then there exists a constant $\mathsf{L}>0$ such that
\begin{enumerate}
\item[(i).] for any $X \in \mathbb{S}_0^n$, $g^k(X)\leq \mathsf{L}I$ for all $k\geq \mathsf{I_o}$;
\item[(ii).] for any $X \in \mathbb{S}_+^n$, $g^{k+1}(X)\leq \mathsf{L}I$ for all $k\geq \mathsf{I_o}$.
\end{enumerate}
\end{lemma}

According to the definitions of $\alpha_j$ and
$\beta_j$, we can further define the sojourn times at the state $1$
and $0$ respectively as follows
\begin{eqnarray*}
\alpha_j^*&\triangleq& \alpha_j-\beta_{j-1}\in\mathbb{N},\\
\beta_j^* &\triangleq& \beta_j-\alpha_{j}\in\{1,\ldots,s\}.
\end{eqnarray*}
The distribution of sojourn times $\alpha_j^*$ and
$\beta_j^*$ is given in the following lemma.
\begin{lemma}[Lemma 4 in~\cite{xiao2009kalman}]\label{lemma:sojourn-time-distribution-lemma}
Denote the joint distribution of $\alpha_j^*$ and $\beta_j^*$ by
$$
\pi(l)\triangleq \mathbb{P}\left(\alpha_1^\ast=a_1,\beta_1^\ast=b_1,\ldots,
\alpha_l^\ast=a_l,\beta_l^\ast=b_l\right),
$$
for any $\alpha_j\in\mathbb{N}$ and $\beta_j\in\{1,\ldots,s\}$.
Then it holds that
\begin{eqnarray*}
\pi(1)&=&\left\{\begin{array}{lcc}
\pi_{b_1}, & \hbox{if~}a_1=1;\\
\pi_0 (\pi_{00})^{a_1-2}\pi_{0b_1}, & \hbox{if~} a_1\geq 2,
\end{array}\right.\\
\pi(l+1)&=&\left\{\begin{array}{lcc}
\pi_{b_lb_{l+1}}\pi(l), & \hbox{if~}a_{l+1}=1;\\
\pi_{b_l0} (\pi_{00})^{a_{l+1}-2}\pi_{0b_{l+1}}, & \hbox{if~} a_{l+1}\geq 2,
\end{array}\right.
\end{eqnarray*}
\end{lemma}

{\it Proof of Theorem~\ref{thm:main-thm}.}
We compute $\mathbb{E}[P_{\beta_1}]$ as follows:
\begin{eqnarray}\label{eqn:P_beta_1}
\mathbb{E}[P_{\beta_{1}}]
&=&\sum_{a_{1}=1}^\infty\sum_{b_{1}=1}^sP_{\beta_{1}}\pi(1)\notag\\
&=&\sum_{b_{1}=1}^s
 \pi_{b_1} h^{b_{1}}\circ g (\Sigma_0)\notag\\
&&+\sum_{a_{1}=2}^{\mathsf{I_o}}\sum_{b_{1}=1}^s\pi_0 (\pi_{00})^{a_1-2}\pi_{0b_1}
h^{b_{1}}\circ g^{a_{1}}(\Sigma_0)\notag\\
&&+\sum_{a_{1}=\mathsf{I_o}+1}^\infty\sum_{b_{1}=1}^s \pi_0 (\pi_{00})^{a_1-2}\pi_{0b_1} h^{b_{1}}\circ g^{a_{1}}(\Sigma_0)\notag\\
&\leq&\sum_{b_{1}=1}^s \pi_{b_1}A^{b_{1}}
(A+K^{(1)}C)\Sigma_0(A+K^{(1)}C)^*
(A^{b_{1}})'\notag\\
&&+\sum_{b_{1}=1}^s A^{b_{1}}\left[
\sum_{a_{1}=2}^{\mathsf{I_o}}\pi_0 (\pi_{00})^{a_1-1}\pi_{0b_1}
(A^{a_1}+K^{(a_1)}C^{(a_1)})\Sigma_0 (\,\cdot\,)^*
\right](A^{b_{1}})'\notag\\
&&+\sum_{b_{1}=1}^s \pi_{b_1}A^{b_{1}}\left(
[A\;\;K^{(1)}]J_1[A\;\;K^{(1)}]^*\right)(A^{b_{1}})'\notag\\
&&+\sum_{b_{1}=1}^s
\sum_{a_{1}=2}^{\mathsf{I_o}}\pi_0 (\pi_{00})^{a_1-2}\pi_{0b_1}
A^{b_{1}}\left([A^{a_1}\;\;K^{(a_1)}]\Sigma_0 [A^{a_1}\;\;K^{(a_1)}]^*\right)(A^{b_{1}})'\notag\\
&&+\sum_{b_{1}=1}^s\left(\sum_{a_{1}=2}^{\mathsf{I_o}}\pi_0 (\pi_{00})^{a_1-2}\pi_{0b_1}
+\pi_{b_1}\right)\sum_{i=0}^{b_1-1}A^iQ(A^i)'\notag\\
&&+\sum_{a_{1}=\mathsf{I_o}+1}^\infty\sum_{b_{1}=1}^s
\pi_0 (\pi_{00})^{a_1-2}\pi_{0b_1} h^{b_{1}}
(\mathsf{L}I)\notag\\
&\triangleq&\Lambda_1+\Lambda_2+\Lambda_3+\Lambda_4+\Lambda_5+\Lambda_6,
\end{eqnarray}
where $J_i\triangleq\left[\begin{array}{ccc}
Q^{(i)}& Q^{(i)}(D^{(i)})'\\
D^{(i)}(Q^{(i)})& D^{(i)}(Q^{(i)})(D^{(i)})'+R^{(i)}
\end{array}\right]$ and the inequality is from Lemmas~\ref{lemma:gx} and~\ref{lemma:riccati>Io}. One can verify that
$\Lambda_3,~\Lambda_4,~\Lambda_5$ and $\Lambda_6$
are all bounded matrices. Then $U\triangleq\Lambda_3+\Lambda_4+\Lambda_5+\Lambda_6$ is also bounded.
 To facilitate the analysis in the following,
we will impose~\eqref{eqn:P_beta_1} to take equality. Without
loss of generality, the conclusions in this paper
still hold for other cases as~\eqref{eqn:P_beta_1} renders us
an upper bound of $\mathbb{E}[P_{\beta_{1}}]$.
Next we vectorize both sides of~\eqref{eqn:P_beta_1}. One has
\begin{eqnarray}\label{eqn:vect-P-beta-1}
\mathbb{E}[\mathrm{vec}(P_{\beta_{1}})]
&=&\sum_{b_{1}=1}^s
(A\otimes A)^{b_1}\Big[
\sum_{a_{1}=2}^{\mathsf{I_o}}\pi_0 (\pi_{00})^{a_1-2}\pi_{0b_1}
(\overline{A^{a_1}+K^{(a_1)}C^{(a_1)}})\otimes
(A^{a_1}+K^{(a_1)}C^{(a_1)})+\notag\\
&&\pi_{b_1}(\overline{A+K^{(1)}C})\otimes
(A+K^{(1)}C)\Big]\mathrm{vec}{(\Sigma_0)}+\mathrm{vec}{(U)}
\end{eqnarray}

Similarly, for any $j\geq 1$, $\mathbb{E}[P_{\beta_{j+1}}]$ can be calculated as
\begin{eqnarray*}
\mathbb{E}[P_{\beta_{j+1}}]&=&\sum_{a_{1}=1}^\infty\sum_{b_{1}=1}^s\cdots\sum_{a_{j+1}=1}^\infty
\sum_{b_{j+1}=1}^s P_{\beta_{j+1}}(l+1)\pi(l+1)\\
&=&\sum_{a_{1}=1}^\infty\sum_{b_{1}=1}^s\cdots\sum_{a_{j+1}=\mathsf{I_o}}^\infty
\sum_{b_{j+1}=1}^s P_{\beta_{j+1}}(l+1)\pi(l+1)\\
&&+\sum_{a_{1}=1}^\infty\sum_{b_{1}=1}^s\cdots\sum_{a_{j+1}=2}^{\mathsf{I_o}-1}
\sum_{b_{j+1}=1}^s \pi_{b_l0} (\pi_{00})^{a_{l+1}-2}\pi_{0b_{l+1}}
h^{b_{l+1}}\circ g^{a_{l+1}}\big(P_{\beta_{j}}(l)\big)\pi(l)\\
&&+\sum_{a_{1}=1}^\infty\sum_{b_{1}=1}^s\cdots\sum_{b_{j}=1}^s
\sum_{b_{j+1}=1}^s \pi_{b_lb_{l+1}} h^{b_{l+1}}\circ g\big(P_{\beta_{j}}(l)\big) \pi(l)\\
&\triangleq& \Gamma_1+\Gamma_2+\Gamma_3.
\end{eqnarray*}
Next we will analyze the boundness of $\Gamma_1, \Gamma_2$ and $\Gamma_3$
one by one.
\begin{eqnarray*}
\Gamma_1&\leq& \sum_{a_{1}=1}^\infty\sum_{b_{1}=1}^s\cdots\sum_{a_{j+1}=\mathsf{I_o}}^\infty
\sum_{b_{j+1}=1}^s h^{b_{l+1}}(\mathsf{L}I)\pi(l+1)\\
&\triangleq & W_1,
\end{eqnarray*}
where the inequality is derived from Lemma~\ref{lemma:riccati>Io} and
$W_1$ is a bounded matrix.
\begin{eqnarray*}
\Gamma_2&\leq& \sum_{a_{1}=1}^\infty\sum_{b_{1}=1}^s\cdots\sum_{a_{j+1}=2}^{\mathsf{I_o}-1}
\sum_{b_{j+1}=1}^s \pi_{b_l0} (\pi_{00})^{a_{l+1}-2}\pi_{0b_{l+1}}
h^{b_{l+1}}\circ \phi_{a_{l+1}}\big(K^{(a_{l+1})},P_{\beta_{j}}(l)\big)\pi(l)\\
&=&\sum_{a_{1}=1}^\infty\sum_{b_{1}=1}^s\cdots\sum_{a_{j+1}=2}^{\mathsf{I_o}-1}
\sum_{b_{j+1}=1}^s \pi_{b_l0} (\pi_{00})^{a_{l+1}-2}\pi_{0b_{l+1}}
A^{b_{l+1}}(A^{a_{l+1}}+K^{(a_{l+1})}C^{a_{l+1}})
P_{\beta_{j}}(l)(\,\cdot\,)^*(A^{b_{l+1}})'\pi(l)\\
&&+\sum_{a_{1}=1}^\infty\sum_{b_{1}=1}^s\cdots\sum_{a_{j+1}=2}^{\mathsf{I_o}-2}
\sum_{b_{j+1}=1}^s \pi_{b_l0} (\pi_{00})^{a_{l+1}-2}\pi_{0b_{l+1}}
\sum_{i=0}^{b_{l+1}-1}A^iQ(A^i)'\pi(l)\\
&&+
\sum_{a_{1}=1}^\infty\sum_{b_{1}=1}^s\cdots\sum_{a_{j+1}=2}^{\mathsf{I_o}-1}
\sum_{b_{j+1}=1}^s \pi_{b_l0} (\pi_{00})^{a_{l+1}-2}\pi_{0b_{l+1}}
A^{b_{l+1}}\left([A^{a_{l+1}}\;K^{(a_{l+1})}]J_{a_{l+1}}
[\,\cdot\,]^*
\right)(A^{b_{l+1}})'\pi(l)\\
&\triangleq& \Gamma_2'+W_2+W_3.
\end{eqnarray*}
It is straightforward to verify that $W_2$ and $W_3$ is bounded.
\begin{eqnarray*}
\Gamma_3&\leq& \sum_{a_{1}=1}^\infty\sum_{b_{1}=1}^s\cdots
\sum_{b_{j+1}=1}^s \pi_{b_lb_{l+1}}
h^{b_{l+1}}\circ \phi_{1}\big(K^{(1)},P_{\beta_{j}}(l)\big)\pi(l)\\
&=&\sum_{a_{1}=1}^\infty\sum_{b_{1}=1}^s\cdots
\sum_{b_{j+1}=1}^s \pi_{b_lb_{l+1}}
A^{b_{l+1}}(A+K^{(1)}C)
P_{\beta_{j}}(l)(\,\cdot\,)^*(A^{b_{l+1}})'\pi(l)\\
&&+\sum_{a_{1}=1}^\infty\sum_{b_{1}=1}^s\cdots
\sum_{b_{j+1}=1}^s \pi_{b_lb_{l+1}}
\sum_{i=0}^{b_{j+1}-1}A^iQ(A^i)'\pi(l)\\
&&+
\sum_{a_{1}=1}^\infty\sum_{b_{1}=1}^s\cdots
\sum_{b_{j+1}=1}^s \pi_{b_l0} \pi_{b_lb_{l+1}}
A^{b_{l+1}}\left([A\;\;K^{(1)}]\,J_{1}\,
[A\;\;K^{(1)}]^*
\right)(A^{b_{l+1}})'\pi(l)\\
&\triangleq & \Gamma_3'+W_4+W_5,
\end{eqnarray*}
where $W_4$ and $W_5$ can be readily shown to be bounded. In summary,
\begin{equation}\label{eqn:P_beta_L}
\mathbb{E}[P_{j+1}]\leq \Gamma_2'+\Gamma_3'+V, ~~~\hbox{for}~j\geq 1,
\end{equation}
where $V\triangleq W_1+W_2+W_3+W_4+W_5$.
By a similar argument, we impose~\eqref{eqn:P_beta_L} to take equality and
take vectorization. From~\eqref{eqn:vect-P-beta-1}~and~\eqref{eqn:P_beta_L},
we can calculate $\mathbb{E}[\mathrm{vec}(P_{\beta_{l+1}})]$
recursively as follows
\begin{equation*}
\mathbb{E}[\mathrm{vec}(P_{\beta_{l+1}})]
= T(H_K)^l\Psi\mathrm{vec}(\Sigma_0)+
T(H_K)^{l-1}\Psi\mathrm{vec}(\Theta_{l-1})+\cdots+
\mathrm{vec}(\Theta_{0}).
\end{equation*}
where $\Theta_{0},\ldots,\Theta_{l-1}$ are the functions of $Q$, $A$, $K^{(i)}$'s and are bounded for
$V$ is bounded,
$$T=[\underbrace{1,\ldots,1}_{s~\hbox{numbers}}]\otimes I_{n^2\times n^2}$$
and $\Psi=[\psi_1^{'},\ldots,\psi_s^{'}]^{'}\in \mathbb{C}^{sn^2\times n^2}$
with
\begin{eqnarray*}\psi_i&=&(A\otimes A)^i\Big[
\sum_{a_{1}=2}^{\mathsf{I_o}}\pi_0 (\pi_{00})^{a_1-2}\pi_{0b_1}
(\overline{A^{a_1}+K^{(a_1)}C^{(a_1)}})\otimes
(A^{a_1}+K^{(a_1)}C^{(a_1)})+\\
&&\pi_{b_1}(\overline{A+K^{(1)}C})\otimes
(A+K^{(1)}C)\Big], ~~~~\hbox{for~}i\in\{1,\ldots,s\}.
\end{eqnarray*}
Therefore, $\mathbb{E}[\mathrm{vec}(P_{\beta_{l+1}})]$ is bounded as
 $l\rightarrow \infty$ if $\rho({H_K})<1$. By some basic algebraic manipulations,  one obtains that $\mathbb{E}\|P_{\beta_{l+1}}\|$ is uniformly bounded if $\rho({H_K})<1$, which completes the proof.
\hfill$\square$

The stability condition in Theorem~\ref{thm:main-thm} is difficult to test.
In the following, we provide an equivalent condition.
In view of this result, Theorem~\ref{thm:main-thm} can be recast as
an LMI feasibility problem. As for the conversion to LMIs using Schur complements, we refer readers to~\cite{boyd1994linear} for details.
\begin{proposition}\label{prop:equal-conditions}
The following statements are equivalent:
\begin{enumerate}
\item[$(i).$]There exists $K\triangleq[K^{(1)},\ldots,K^{(\mathsf{I_o}-1)}]$, where
$K^{(i)}$'s are matrices with compatible dimensions, such that
$\rho({H_K})<1$, where $H_K$ is defined in~\eqref{def:H-k};

\item[$(ii).$]There exist $X_1>0,\ldots,X_s>0$ and $K_1,\ldots,K_{\mathsf{I_o}-1}$
such that
\begin{eqnarray}
&&\sum_{i=1}^s\pi_{i0}\pi_{0j}\sum_{l=2}^{\mathsf{I_o}-1}
(\pi_{00})^{l-2}A^j(A^l+K_lC^{(l)})X_i(A^l+K_lC^{(l)})^*(A^j)'\notag\\
&&\hspace{1.5cm}+\sum_{i=1}^s\pi_{ij}A^j(A+K_1C)X_i(A+K_1C)^*(A^j)'<X_j,~~~
\hbox{for~all~} j\in\mathcal{S}/\{0\}.
\end{eqnarray}
\end{enumerate}
\end{proposition}
{\it Proof.}
$(i)\Rightarrow (ii)$~Since $\rho(H_K)<1$, we have
\begin{equation}\label{eqn:H_k-serial}
(I-H_K)^{-1}=I+H_K+(H_K)^2+\cdots.
\end{equation}
We define a linear space $\mathbb{H}_+^n$ as
$$\mathbb{H}_+\triangleq \{[H_1,\ldots,H_s]:H_j\in\mathbb{S}_+^n,~\forall~j\in\mathcal{S}\}.$$
Then define a norm on $\mathbb{H}_+^n$ as
$$\|H\|_\ast\triangleq\sum_{i=1}^s\|H_i\|$$
for any $H=[H_1,\ldots,H_s]\in\mathbb{H}_+^n$.
For any $G\triangleq [G_1,\ldots,G_s],H\triangleq [H_1,\ldots,H_s]\in\mathbb{H}_+^n$, we say that $G\succeq H$ (and $G\succ H$)
if $G_i\geq H_i$ (and $G_i>H_i$) for all $i\in\{1,\ldots,s\}$.

Since
$(I-H_K)$ is nonsingular and  vectorization is a bijective
mapping, for any $H\in\mathbb{H}_+^s$, there exists a unique
matrix $X\triangleq [X_1,\ldots,X_s]\in \mathbb{C}^{n\times sn}$ such that
\begin{equation}\label{eqn:H-X-eqn}
\mathrm{vec}(H)=(I-H_K)\mathrm{vec}(X).
\end{equation}
In what follows, we shall show $X\in\mathbb{H}^{n}_+$.
From~\eqref{eqn:H_k-serial}, we have
\begin{eqnarray*}
\mathrm{vec}(X)&=&(I-H_K)^{-1}\mathrm{vec}(H)\\
&=&\sum_{i=0}^\infty (H_K)^i\mathrm{vec}(H).
\end{eqnarray*}
Taking the inverse mapping of vectorization gives
$X\succeq H$, implying $X\in\mathbb{H}^{n}_+$.
Similarly, by taking the inverse mapping of vectorization
on~\eqref{eqn:H-X-eqn}, we have
\begin{eqnarray*}
H_j&=&X_j-\sum_{i=1}^s\pi_{i0}\pi_{0j}\sum_{l=2}^{\mathsf{I_o}-1}
(\pi_{00})^{l-2}A^j(A^l+K^{(l)}C^{(l)})X_i(A^l+K^{(l)}C^{(l)})^*(A^j)'\notag\\
&&\hspace{0.5cm}-\sum_{i=1}^s\pi_{ij}A^j(A+K^{(1)}C)X_i(A+K^{(1)}C)^*(A^j)',
~~~\hbox{for~all~} j\in\mathcal{S}/\{0\}.
\end{eqnarray*}
where this claim follows as asserted.

$(ii)\Rightarrow (i)$~Define an operator $\mathcal{L}_K\triangleq (\mathcal{L}_{K,1},\ldots,\mathcal{L}_{K,s}):~\mathbb{H}_+^n\rightarrow
\mathbb{H}_+^n$ as
\begin{eqnarray}
\mathcal{L}_{K,j}(H)&\triangleq &\sum_{i=1}^s\pi_{i0}\pi_{0j}\sum_{l=2}^{\mathsf{I_o}-1}
(\pi_{00})^{l-2}A^j(A^l+K_lC^{(l)})H_i(A^l+K_lC^{(l)})^*(A^j)'\notag\\
&&\hspace{1cm}+\sum_{i=1}^s\pi_{ij}A^j(A+K_1C)H_i(A+K_1C)^*(A^j)',.
\end{eqnarray}
where $K=[K_1,\ldots,K_{\mathsf{I_o}-1}]$, and $H=[H_1,\ldots,H_s]\in\mathbb{H}_+^n$.
It is evident that $\mathcal{L}_K(\alpha H)=\alpha \mathcal{L}_K(H)$ for
any $\alpha\in\mathbb{R}$, and that
$\mathcal{L}_K( G)\succeq  \mathcal{L}_K(H)$ for
$G\succeq H$.
From the hypothesis of $(ii)$, we conclude that there
exists a $\mu\in(0,1)$ such that
$\mathcal{L}_K(X)\preceq \mu X$, where $X\triangleq [X_1,\ldots,X_s]$.
In addition, for any given $H_0\in\mathbb{H}_+^n$,
there always exists an $r>0$ such that
$H_0\preceq rX$. Therefore, for $k\in\mathbb{N}$,
$$
\mathcal{L}_K^k(H_0)\preceq r\mathcal{L}_K^k(X)\preceq r\mu\mathcal{L}_K^{k-1}(X)
\preceq \cdots \preceq r\mu^{k} X,
$$
which leads to $\|\mathcal{L}_K^k(H_0)\|_\ast\leq r\mu^k\|X\|_\ast$.
As $k\rightarrow \infty$, we have
$\lim_{k\rightarrow \infty}\|\mathcal{L}_K^k(H_0)\|_\ast=0$.
Note that $\mathrm{vec}\left(\mathcal{L}_K(H_0)\right)=
H_K\mathrm{vec}(H_0)$.
Combining all the above observations, we have
$$\lim_{k\rightarrow \infty}(H_K)^k \mathrm{vec}(H_0)=0,$$
which implies $\rho(H_K)<1$. This completes the proof.
\hfill$\square$

\section{Comparison with~\cite{xiao2009kalman}}
\label{section:comparison}
In this part, we compare our result with those in~\cite{xiao2009kalman} and show the advantages of ours. Recall that the sufficient condition  in~\cite{xiao2009kalman} is
$\rho(\Phi)<1$ where
$$\Phi\triangleq \left[d_1^{(1)}\mathbf{P}+\sum_{l=2}^{\mathsf{I_o}-1}(\pi_{00})^{l-1}d_l^{(1)}
\mathbf{Q}\right]\mathrm{diag}\left(\|A\|^2,\ldots,\|A^s\|^2
\right),
$$
with $\mathbf{P},~\mathbf{Q}$ being defined in~\eqref{def:H-k} and  $d_l^{(1)}\triangleq\min_{K^{(l)}}\|A^{(l)}+K^{(l)}C^{(l)}\|^2$.
\subsection{Invariance with Respect to Similarity Transformations}
Theoretically, a state variable transformation (i.e., a similarity transformation from a linear system $(A,B,C,D)$ to $(S^{-1}AS,S^{-1}B,CS,D)$
through the nonsingular matrix $S$ does not change the stability
considered in this work. However, different state variable transformations
may generate opposite conclusions from the stability condition
given in~\cite{xiao2009kalman}. The invariance of stability behavior with respect to state variable transformations can be reflected well from the stability conditions presented by this work.
\begin{proposition}
Let $S\in\mathbb{C}^{n\times n}$ be nonsingular.
Suppose there exists $K\triangleq[K^{(1)},\ldots,K^{(\mathsf{I_o}-1)}]$, where
$K^{(i)}$'s are matrices with compatible dimensions, such that
$\rho({H_K})<1$, where ${H_K}$ is defined in~\eqref{def:H-k} for
$(A,C)$. Then, there always exists
$\tilde K\triangleq S^{-1}[K^{(1)},\ldots,K^{(\mathsf{I_o}-1)}]$
such that $\rho({\tilde H_{\tilde K}})<1$, where ${\tilde H_{\tilde K}}$ is defined for
$(\tilde A, \tilde C)\triangleq (S^{-1}AS, CS)$ in accordance with~\eqref{def:H-k}.
\end{proposition}
The proof follows from~Proposition~\ref{prop:equal-conditions} and
direct calculation.
 We use the following example to illustrate this idea.
\begin{example}\label{example:1}
Consider the system
$$
A=\left[\begin{array}{ccc}
1.3 & 0.3 \\
0 & 1.2
\end{array}\right],~~
C=[\,1~~1\,],
$$
$Q=I_{2\times 2}$ and $R=1$, and the bounded Markovian packet-loss process with
transition probability matrix given by
\begin{equation}\label{def:trans-matrix-1}
\mathbf{\Pi}=\left[\begin{array}{cccc}
0.6 & 0.2 & 0.2 \\
0.8 & 0.1 & 0.1 \\
0.8 & 0.1 & 0.1
\end{array}\right].
\end{equation}
From~\cite{xiao2009kalman}, we have $d_1^{(1)}=1.2200$ and $\rho(\Phi)=0.7352<1$.
Let
$$S=\left[\begin{array}{ccc}
1 & 5 \\
0 & 1
\end{array}\right].$$
For the system $(\tilde A, \tilde C)\triangleq (S^{-1}AS, CS)$,
we have $\tilde d_1^{(1)}=1.3632$ and $\rho(\tilde \Phi)=1.5202>1.$
\end{example}

\subsection{Conservativity Comparison}
The stability condition given in this work is less conservative compared with that in~\cite{xiao2009kalman}, since the latter condition implies the former one. To show this, we need the following proposition.
\begin{proposition}\label{prop:less-conservative}
Define
$$\Phi_K\triangleq \left[d_1\mathbf{P}+\sum_{l=2}^{\mathsf{I_o}-1}(\pi_{00})^{l-1}d_l
\mathbf{Q}\right]\mathrm{diag}\left(\|A\|^2,\ldots,\|A^s\|^2
\right),
$$
where
$\mathbf{P}$ and $\mathbf{Q}$ are defined in~\eqref{def:H-k} and
$d_l\triangleq\|A^{(l)}+K^{(l)}C^{(l)}\|^2$, and
$K\triangleq[K^{(1)},\ldots,K^{(\mathsf{I_o}-1)}]$ with
$K^{(l)}$'s of compatible dimensions.
If there exists $K$ such that
$\rho({\Phi_K})<1$, then $\rho({H_K})<1$.
\end{proposition}
{\it Proof.}
If treating a scalar as the Kronecker product of two other scalars,
similar to Proposition~\ref{prop:equal-conditions}, we can claim
that, if and only if $\rho({\Phi_K}')<1$, there exists a vector $$x\triangleq [x_1,\ldots,x_s],$$ where $x_j>0\hbox{~for~all~} j\in\mathcal{S}/\{0\},$ such that
\begin{equation*}
\sum_{i=1}^s\pi_{i0}\pi_{0j}\sum_{l=2}^{\mathsf{I_o}-1}
(\pi_{00})^{l-2}d_l\|A^j\|^2x_i+\sum_{i=1}^s\pi_{ij}d_1\|A^j\|^2x_i
<x_j.
\end{equation*}
The submultiplicativity and subadditivity of a matrix norm result in the following inequality
\begin{eqnarray}\label{eqn:Psi-2-LMI}
&&\Big\|\sum_{i=1}^s\pi_{i0}\pi_{0j}\sum_{l=2}^{\mathsf{I_o}-1}
(\pi_{00})^{l-1} x_i A^j(A^l+K^{(l)} C^{(l)})(A^l+K^{(l)} C^{(l)})^*(A^j)'\notag\\
&&\hspace{1.5cm}+\sum_{i=1}^s\pi_{ij}x_i
A^j(A+K^{(1)} C)(A+K^{(1)} C)^*(A^j)'\Big\|<x_j,~~~\hbox{for~all~} j\in\mathcal{S}/\{0\}.
\end{eqnarray}
Let $X_j=x_j I_{n\times n}$. Then we obtain from~\eqref{eqn:Psi-2-LMI} that
\begin{eqnarray*}
&&\sum_{i=1}^s\pi_{i0}\pi_{0j}\sum_{l=2}^{\mathsf{I_o}-1}
(\pi_{00})^{l-2}A^j(A^l+K^{(l)}  C^{(l)})X_j(A^l+K^{(l)}  C^{(l)})^*(A^j)'\notag\\
&&\hspace{1.5cm}+\sum_{i=1}^s\pi_{ij}A^j(A+K^{(1)}  C)X_i(A+K^{(1)}  C)^*(A^j)'<X_j.
\end{eqnarray*}
Therefore $\rho(H_{K})<1$, which completes the proof.
\hfill$\square$

In virtue of Proposition~\ref{prop:less-conservative}, it is evident that $\rho(\Phi)<1$ implies $\rho(H_{K^\star})<1$,
where $K^\star\triangleq [K_1^\star,\ldots,K_{\mathsf{I_o}-1}^\star]$ with
$K_i^\star \triangleq \arg\min_{K^{(i)}}\|A^{(i)}+K^{(i)}C^{(i)}\|^2$.


\addtocounter{example}{-1}
\renewcommand{\theexample}{\arabic{example}$\mathbf{~(cont'd)}$}
\begin{example}
We continue to consider Example~\ref{example:1} with
 an alternative
transition probability matrix
$${\mathbf \Pi}_1=
\left[\begin{array}{cccc}
0.6 & 0.2 & 0.2\\
0.6 & 0.2 & 0.2\\
0.6 & 0.2 & 0.2
\end{array}\right].$$
From Theorem 2 in~\cite{xiao2009kalman}, we obtain
$\rho(\Phi)=1.4704>1$.
By solving an LMI feasibility problem using the $\mathbf{cvx}$ in Matlab, we see that our Theorem~\ref{thm:main-thm} still holds with
a group of feasible variables
$$X_1=X_2=\left[\begin{array}{ccc}
0.1081   & 0.0243\\
    0.0243  &  0.1042
\end{array}\right]~\hbox{and}~K=
\left[\begin{array}{c}
   -0.8079\\
   -0.5914
\end{array}\right].$$

If we consider the transition probability matrix which only
allows the maximum length of consecutive packet losses to be $1$, i.e.,
$$
\mathbf{\Pi}_2=\left[\begin{array}{ccc}
0.6   & 0.4\\
0.8  &  0.2
\end{array}\right],
$$
then $\rho(\Phi)=0.49<1$ and the condition
in our Theorem~\ref{thm:main-thm} holds.
When we increase $\pi_{11}$ in $\mathbf{\Pi}_2$ from
$0.2$ to $0.5$, one can verify numerically that
$\rho(\Phi)>1$ while Theorem~\ref{thm:main-thm} of this paper still holds.
\end{example}


\section{Conclusion}
We have considered the bounded Markovian packet-loss process model and the notion of the peak-covariance stability for the Kalman filtering.
A sufficient stability condition with bounded Markovian packet losses was established. Different from that of~\cite{xiao2009kalman}, this stability check can be recast as an LMI feasibility problem.
Then we compared the proposed condition with that of~\cite{xiao2009kalman},
showing that our condition prevails from at least two aspects:
\begin{inparaenum}[\itshape 1\upshape)]
\item Our stability condition is invariant with respect to similarity state  transformations, while the previous result is not;
\item More importantly, our condition is proved to be less conservative than the previous one.
\end{inparaenum}
Numerical examples
were provided to demonstrate the effectiveness our result compared with the literature.

\bibliographystyle{IEEETran}
\bibliography{sj_reference,sj_reference1,sj_reference2}

\medskip

\medskip

\medskip

\noindent {\sc Junfeng Wu and Karl H. Johansson} \\
{\noindent {\small  ACCESS Linnaeus Centre,
   School of Electrical Engineering,
\\
KTH Royal Institute of Technology, Stockholm 100 44, Sweden }\\}
       {\small Email: {\tt\small junfengw@kth.se, kallej@kth.se}

\medskip

{\noindent {\sc Ling Shi}} \\
{\noindent {\small Department of Electronic and Computer Engineering,\\ Hong Kong University of Science and Technology, Hong Kong}}\\  {\small Email: } {\tt\small eesling@ust.hk}

{\noindent {\sc Lihua Xie}} \\
{\noindent {\small School of Electrical and Electronic Engineering,\\  Nanyang Technological University, Singapore}}\\  {\small Email: } {\tt\small elhxie@ntu.edu.sg}

\end{document}